\documentstyle[12pt]{article}

\newcommand{\beq}{\begin{eqnarray}}
\newcommand{\eeq}{\end{eqnarray}}

\newcommand{\hal}[1]{{#1 \over 2}}
\newcommand{\cket}[1]{| #1 \rangle}
\newcommand{\bra}[1]{\langle #1 |}

\newcommand{\pslash}{{p\hspace{-5pt}/}}

\newcommand{\gpNN}{g_{\pi NN}}

\newcommand{\gpNR}{g_{\pi NN^*}}
\newcommand{\geNR}{g_{\eta NN^*}}
\newcommand{\gpRR}{g_{\pi N^* N^*}}

\begin{document}

\begin{center}
{\large Suppression of $\pi NN^*$ Coupling
and Chiral Symmetry\\ }
\vspace*{0.5cm}
 { D. Jido\footnote{e-mail address: jido@th.phys.titech.ac.jp} 
 and M. Oka \\
 Department of Physics, Tokyo Institute of Technology\\
 Meguro, Tokyo 152  Japan}

\vspace*{0.5cm}
{A. Hosaka \\
Numazu College of Technology \\
3600 Ooka, Numazu 410 Japan}
\end{center}
\vspace*{1cm}
\abstract{
Meson-baryon couplings between positive and negative parity baryons are 
investigated using two point correlation functions 
in the soft meson limit. 
We find that the $\pi NN^*$ coupling vanishes due to chiral symmetry, 
while the $\eta NN^*$ coupling remains finite.
We perform an analysis based on the algebraic method for
$SU(2)$ and $SU(3)$ chiral symmetry, and find that baryon axial charges 
play an essential role for vanishing coupling constants. }

\newpage

Baryon resonance is a suitable laboratory to test 
effective models of QCD at low energy.
Not only masses but also transition matrix elements are
subject to recent research interests. 
New data will soon become available
from facilities such as TJNAL (former CEBAF), COSY, ELSA, MAMI
and possibly from SPring8~\cite{INT}.
In particular, transition form factors are useful to investigate 
details of wave functions.

Among various baryon resonances, negative parity resonances $N^*$
have particularly interesting properties.
For example, $N(1535)$  has relatively large branching 
ratio for the decay $N(1535) \to \eta N$ which is comparative to that 
of $N(1535) \to \pi N$~\cite{PDG}.  
Considering the difference in the available phase spaces,
this leads to a relatively
large coupling constant for $\eta NN(1535)$.
One may also look at the problem in the following way.  
Using the experimental decay widths of the resonance, we obtain
$\gpNR \sim 0.7$ and $\geNR \sim 2$.  
These values, in particular $\gpNR$, 
are much smaller than, for example, $\gpNN \sim 13$.  
Thus, one may ask 
why especially the coupling $\gpNR$ is so suppressed 
as compared with $\gpNN$.

Previously, Jido, Kodama and Oka~\cite{jko}
have studied masses  of 
negative parity baryons $B^*$ in the QCD sum
rule~\cite{svz,rry}.
A method to extract the information of $B^*$ from a 
correlation function has been formulated.
The resulting masses were generally in good agreement with data.  
One of the motivations in the previous work
was to study chiral properties of the negative parity baryons.
In fact, it was shown that the mass splitting between a pair  
of positive and 
negative parity baryons is caused by nonzero quark condensate,  
suggesting that they form a parity doublet.  
The purpose of the present work is then to look for other 
quantities which are governed by chiral properties of 
baryons.  
Indeed, we will see that meson-baryon couplings are such 
examples.  

Our starting point is the observation that baryon interpolating 
fields 
couple to both positive and negative parity baryons~\cite{cdks,i}.
Let an interpolating field $J$ couple to a positive parity nucleon $N$:
\beq
\label{JNtopos}
\langle 0 | J | N \rangle 
= \lambda_{N} u_{N} \, ,
\eeq
where $\lambda_{N}$ is the strength of the coupling and $u_{N}$
is the spinor for the nucleon.  
Then the same interpolating field $J$ couples to negative parity 
nucleons $N^*$ also:
\beq
\label{JNtoneg}
\langle 0 | J | N^* \rangle 
= \lambda_{N^*} \gamma_5 u_{N^*} \, ,
\eeq
In general, one might have chosen different interpolating fields 
for $N$ and $N^*$.  
However, by adopting the same type of 
interpolating field, $N$ and $N^*$ could   
be regarded as a parity doublet~\cite{jko}.  
This is our view point in the present work.  

The general nucleon interpolating field without derivatives 
is given by the superposition of two independent terms~\cite{ept} (see 
also (\ref{dectwo}) and discussions that follow):
\begin{eqnarray}
   J(x;t) & = & \varepsilon^{abc} [(u_{a}(x)Cd_{b}(x))
   \gamma_{5} u_{c}(x) + t (u_{a}(x) C \gamma_{5} d_{b}(x))
   u_{c}(x)] \, ,
        \label{eq:nucur}
\end{eqnarray}
where  $a$, $b$ and $c$ are color indices, $C = i \gamma_{2} \gamma_{0}$
(in the standard notation) is the charge conjugation matrix, and
$t$ is the mixing parameter of the two terms.
It is known that the Ioffe's current $J(x;t=-1)$~\cite{i} 
couples strongly to the positive parity nucleon~\cite{ept}, while
it was found in Ref.~\cite{jko}
that  by choosing $t \sim 0.8$, $J(x;t=0.8)$ is made to be optimized 
for negative parity nucleons.  
The role of $t$ is not only to control the coupling strength $\lambda$, 
but also to change chiral structure of $J$.  
In general the states which $J$ couples to
are not always physically  observed ones;  
they could be some linear combinations of them.    
However, 
the previous sum rule study for masses of negative parity baryons  
implies that the current adopted there ($t \sim 0.8$) 
couples strongly to the lowest negative parity baryons including 
$N(1535)$.  

Now let us investigate meson-baryon coupling constants.   
First we consider the $\pi NN^*$ coupling.  
We follow the 
method used by Shiomi and Hatsuda~\cite{sh} and study
the two point correlation function between
the vacuum and a one pion state.
In the soft pion limit 
$(q^{\mu}\to 0)$, the correlation function is given by: 
\beq
   \Pi^{\pi}(p) &=& i \int d^4 x \, e^{ip\cdot x}
        \bra{0} TJ(x;s) \bar{J}(0;t) \cket{\pi (q=0)} \nonumber \\
        &=&
        i  (\Pi_{0}^{\pi}(p^{2})\gamma_{5}  + \Pi_{1}^{\pi}(p^{2}) 
            \pslash\gamma_{5} ) \, ,
   \label{eq:cor}
\eeq
where $J(x;t)$ or $J(x;s)$ is defined in (\ref{eq:nucur}). 
Here the Lorentz structure of the two terms of (\ref{eq:cor}) is  
determined by the total parity of the two point function.  
In general, the correlation function (\ref{eq:cor}) contains 
contributions from $\gpNN$, $\gpNR$ and $\gpRR$.  
It turns out that
the information of the $\pi NN^*$ coupling is involved in
the $\pslash \gamma_5$ term of (\ref{eq:cor}).  
The reason is the following.  
Using the phenomenological $\pi NN^*$ lagrangian
\begin{equation}
    \label{eq:defpNN}
    {\cal L}_{\pi NN^*} = \gpNR \bar{N}^{*} \tau^{a} \pi_{a} N + 
    (h.c.),
\end{equation}
with standard notations, 
the $\pi NN^{*}$ contribution in the
$\Pi^{\pi}(p)$ is given in the soft pion limit by
\begin{equation}
   \label{eq:pNN*}
    \gpNR \lambda_{N}(t) \lambda_{N^{*}}(s)
    \left[{p^{2} + m_{N}m_{N^{*}} \over (p^{2} - m_{N}^{2}) (p^{2} -
    m_{N^{*}}^2)} + { \pslash (m_{N} + m_{N^{*}}) \over (p^{2} -
    m_{N}^{2}) (p^{2} - m_{N^{*}}^2)}\right]\gamma_{5} \, .
\end{equation}
Here we have picked up the term where $J(t)$ couples to $N$ and $J(s)$ 
to $N^*$, corresponding to the physical process $N^* \to N + \pi$.  
Thus, the $\pi NN^*$ coupling is contained in both $\gamma_{5}$ and  
$\pslash \gamma_{5}$ terms.
In contrast, the $\pi NN$ (also and $\pi N^* N^*$) 
coupling gives only the $\gamma_5$ term:
\begin{equation}
   \gpNN \lambda_{N}(t) \lambda_{N}(s) {i \gamma_{5} \over
    p^{2} - m_{N}^{\; 2}} \, ,    \label{eq:pheNN}
\end{equation}
as derived from the $\pi NN$ interaction lagrangian
\begin{equation}
    {\cal L}_{\pi NN} = \gpNN \bar{N} i \gamma_{5} \tau^{i} \pi^{i}
    N\, .
\end{equation}
Therefore, the $\pslash \gamma_5$ term involves information only of 
the $\pi NN^*$ coupling.

In recent reports, we have computed the two point correlation 
function (\ref{eq:cor}) in the operator product expansion (OPE) 
and found 
that it vanishes up to order dimension eight~\cite{jho}.  
We have demonstrated that this result is a consequence of
chiral symmetry of the interpolating field $J(x)$.  
Recently, Cohen and Ji have classified various 
hadron interpolating fields based on chiral 
symmetry~\cite{coji}.  
Then they were able to make several model independent predictions.  
From this point of view, our finding of vanishing
coupling constants is also one of such examples.  

Let us briefly repeat how the term relevant to $\gpNR$ vanishes.  
First we rewrite the one pion matrix element (\ref{eq:cor})
in terms of the commutation relation with the isovector axial 
charge $Q_{A}^{a}$:
\begin{eqnarray}
\Pi^{\pi^a} (p) &=& \lim_{q \to 0}  
\int d^4x e^{ipx} \bra{0} T J(x) \bar J(0) \cket{\pi^a(q)} 
\nonumber \\
&=& - \frac{i}{\sqrt{2}f_\pi} \int d^4x e^{ipx} \bra{0} [ Q_A^a , 
T J(x) \bar J(0) ] \cket{0} \nonumber \\
&=& - \frac{i}{2\sqrt{2}f_\pi} 
\int d^4x e^{ipx} 
\{ \gamma_5 \tau^a, \bra{0} T J(x) \bar J(0)  \cket{0} \}  \, .
\label{Q5comm}
\end{eqnarray}
Here we have used the transformation property of the interpolating 
field $J$:
\begin{equation}
\label{transJ}
[ Q_A^a , J] =  \frac{1}{2} \gamma_5 \tau^a J \, . 
\end{equation}
Using the Lorentz structure of the vacuum to vacuum matrix element 
of (\ref{Q5comm}):
\beq
\int d^4 x e^{ipx} \bra{0}  J(x) \bar J(0)  \cket{0} \sim A \pslash + B
1, 
\eeq
the $\pslash \gamma_5$ term disappears in (\ref{Q5comm}).
This  is 
the basis for the vanishing coupling constant for $\gpNR$.  

We should make one remark here.  
When writing the phenomenological correlation function 
(\ref{eq:pNN*}), only one term  for 
$N^* \to N + \pi$ has been considered, whereas in the 
theoretical expression (\ref{eq:cor}) 
another contribution from the reversed process 
$N + \pi \to N^*$  is also contained.  
If the both contributions are included, the phenomenological 
expression for the $\pslash \gamma_5$ term is factored out by 
$\lambda_{N}(t) \lambda_{N^{*}}(s) - 
 \lambda_{N}(s) \lambda_{N^{*}}(t)$, which vanishes  
when $s = t$.   
In fact, in the OPE, the $\pslash \gamma_5$ term  of 
the correlation function 
has the common factor $s-t$~\cite{jho}.  
This means that when $s = t$, 
the $\pslash \gamma_5$ term 
vanishes due to this trivial factor.  
However, in Eq. (\ref{Q5comm}) that term  
has dropped out not by this factor 
but due to chiral symmetry of the interpolating field.   
This applies to the whole discussions of this paper, 
and conclusions we draw are not disturbed by this trivial 
factor.

Now the crucial point in deriving the vanishing $\gpNR$ is that the 
baryon interpolating field $J$ transforms as the fundamental 
representation of the chiral group $SU(2)_{R} \times SU(2)_{L}$,   
independent of the choice of the parameter $t$ or $s$.  
To look at this point in some more detail, let us investigate 
the algebraic structure of the interpolating field.  
A baryon interpolating field consists of three right and left 
quarks.  
The representation is decomposed as
\beq
\left[ \left( \textstyle{\hal{1}},0 \right) 
           + \left( 0, \textstyle{\hal{1}} \right) \right] ^{3}
&= &
\left[ \left( \textstyle{\hal{3}},0 \right) 
           + \left( 0, \textstyle{\hal{3}} \right) \right]
+ 3 \, \left[ \left( \textstyle{\hal{1}},1 \right) 
           + \left( 1, \textstyle{\hal{1}} \right) \right] \nonumber \\
& & 
\label{dectwo}
+ 3 \, \left[ \left( \textstyle{\hal{1}},\tilde 0 \right) 
           + \left( \tilde 0, \textstyle{\hal{1}} \right) \right]
+ 2 \, \left[ \left( \textstyle{\hal{\tilde 1}},0 \right) 
           + \left( 0, \textstyle{\hal{\tilde 1}} \right) \right] \, ,
\eeq
where, according to Ref.~\cite{coji}, tildes imply that a pair of left
or 
right quarks are coupled to the isoscalar singlet.  
The two relevant terms for the nucleon are then 
$( \hal{1}, \tilde 0) + ( \tilde 0, \hal{1} )$ and 
$( \hal{\tilde 1}, 0) + ( 0, \hal{\tilde 1})$.  
The essential point here is that there is no distinction between the 
two from the algebraic point of view as they both belong to 
the same fundamental representation of the chiral group of 
$SU(2)_R \times SU(2)_L$.  
In other words, the two interpolating fields carry the same $SU(2)$ 
axial charge.  
Therefore, for all $t$ (or $s$), the interpolating field 
(\ref{eq:nucur}) 
transforms in the same way as (\ref{transJ}), and hence the
$\pslash \gamma_5$ term vanishes.

Now we discuss the $\eta N N^*$ coupling.  
Here, within $SU(2)$, 
the $\eta$ meson may be regarded as an isospin singlet.  
The discussion for the pion is extended by, for example, simply 
replacing the isospin Pauli matrices $\tau^a$ by the unit matrix 
$1$, and now the $U(1)_A$ property (actually charge) 
of the current $J$ becomes relevant.   
The crucial observation is that  
while $J$ carries a definite $SU(2)$ axial charge, it does not 
carries a definite $U(1)$ axial charge.  
This is verified through the observation that 
$( \hal{1}, \tilde 0)$ consists of one right quark and two left 
quarks, while
$( \hal{\tilde 1}, 0)$ consists of 
three right quarks.  
Because of this, the $U(1)_A$ transformation rule does not respect 
any symmetry relation such as (\ref{transJ}), and therefore, 
$\geNR$ no longer vanishes.  

To summarize briefly, we have seen that symmetry properties of the 
interpolating field $J$ lead to $\gpNR = 0$  while 
$\geNR \neq 0$.  
Phenomenologically, these properties seem to be well satisfied by 
the negative parity resonance $N(1535)$, suggesting that the 
properties of the resonance are strongly governed by chiral symmetry.  

The situation becomes less trivial for the three flavor case of 
$SU(3)_{R} \times SU(3)_{L}$.  
The reason is that while the $SU(3)$ baryons belong to an octet 
representation of the diagonal vector group $SU(3)_V$, their behavior 
under axial transformations is not uniquely determined.  
But once again, for nonstrange nucleons, we find that 
the $\pslash \gamma_{5}$ term disappears.  

The decomposition of baryon interpolating fields composed of
three quarks under $SU(3)_R \times SU(3)_L$ is
accomplished as
\beq
\left[ (3,1) + (1,3) \right]^3
&=&
\left[ (10,1)+(1,10) \right] +  3 \; \left[ (6,3)+(3,6) \right]
\nonumber \\
\label{decthree}
&+& \; 3 \; \left[ (3, \bar 3) + (\bar 3, 3)  \right]
+ 2 \; \left[ (8,1) + (1,8) \right]
+ \left[ (\tilde 1, 1) + (1, \tilde 1) \right] \, .
\eeq
Here the multiplets assigned to the spin 1/2 octet baryons are
$(8,1) + (1,8)$
and
$ (3, \bar 3) + (\bar 3, 3) $, as they both
transform as an octet under the $SU(3)_V$ transformation.
In contrast, they transform differently under $SU(3)$ axial 
transformations. 
Explicitly, denoting the baryon interpolating field which belongs 
to the $SU(3)$ multiplet $(p, q)$ by
$J_a^{(p,q)} (a=1,2, \cdots 8) $,
the transformation rule is
\beq
\label{QAa}
[Q_A^a, J^{(8,1)}_b] &=&  i f_{abc} J^{(8,1)}_c \, , \\
\label{QAb}
[Q_A^a, J^{(1,8)}_b] &=& - i f_{abc} J^{(1,8)}_c \, , \\
\label{QAc}
[Q_A^a, J^{( 3, \bar3)}_b] &=& d_{abc} J^{(3, \bar 3)}_c \, , \\
\label{QAd}
[Q_A^a, J^{(\bar 3, 3)}_b] &=& - d_{abc} J^{(\bar 3, 3)}_c \, ,
\eeq
where $f_{abc}$ and $d_{abc}$ are the structure constants 
of $SU(3)$.

Consider, for example, the transition $p^* \to p \pi^0$.
The interpolating field is a superposition of
$(8,1)+(1,8)$ and $(3, \bar 3) + (\bar 3, 3)$
with the parameter $\beta (= {1-t \over 1+t})$:
\beq
\label{J38}
J_a (\beta) &=& J^8_a + \beta J^3_a \,  \\
  &J^8_a & \; \equiv \; J^{(8,1)}_a + J^{(1,8)}_a \,  \\
  &J^3_a & \; \equiv \; J^{(\bar 3,3)}_a + J^{(3, \bar 3)}_a \, .
\eeq
Thus the correlation function takes the form
\beq
\Pi_{ab} = \langle J_a (\alpha), \bar J_b (\beta) \rangle
    = \langle J^8_a + \alpha J^3_a, \bar J^8_b + \beta \bar J^3_b
\rangle \, ,
\eeq
Since the flavor structure of the proton is
$p \sim \lambda^4 + i \lambda^5$, we investigate the response
of the 44, 45, 54 and 55 components of the two point functions
under $Q_A^{a=3}$. After some computation, we find that
\beq
\, [Q_A^3, \Pi_{pp}] &\sim& [Q_A^3, \Pi_{4+i5, 4+i5}] \nonumber \\
&=& [Q_A^3, \Pi_{44} + i\Pi_{54} -i \Pi_{45} + \Pi_{55}] \nonumber \\
\label{Pipn}
&=& \hal{1} \left(
\{ i \gamma_5 , \Pi^{(88)}_{pp} \}
+ \alpha \{ i \gamma_5 , \Pi^{(83)}_{pp} \} \right. \nonumber \\
& & + \left. \; \beta \{ i \gamma_5 , \Pi^{(38)}_{pp} \}
+ \alpha \beta \{ i \gamma_5 , \Pi^{(33)}_{pp} \} \right) \, .
\eeq
A crucial point here is that the result is written only in terms of the
anticommutators with $\gamma_5$.
From the 
algebraic point of view, this result for the proton and neutron
follows completely from the symmetry property of 
$(p,n)$ under $SU(2)_R \times SU(2)_L$ transformations.  
Indeed similar arguments in the preceding paragraphs can apply 
here by identifying, for example, $(8,1)$ with $(\tilde{\frac{1}{2}},
0)$, 
and 
$(3, \bar 3)$ with $(\frac{1}{2}, \tilde 0)$.
In other words, both $J^8_N$ and $J^3_N$ ($N = p, n$) 
carry the same $SU(2)$ axial charge.  
Therefore, 
the anticommutation relation (\ref{transJ}) follows from 
(\ref{QAa}) -- (\ref{QAd}) for $a = 1,2,3$.  
On the other hand, the couplings of $\eta$ (a combination of 
$\eta_{1}$ and $\eta_{8}$) involve the axial charge $Q_{A}^{8}$ in the 
case of $SU(3)$. The $\eta_{8} NN^{*}$ coupling does not vanish, 
since the axial charges $Q_{A}^{8}$ of $J^8_N$ and $J^3_N$ 
are not the same.  
The singlet part for the $\eta_1 NN^*$ coupling 
follows the argument as given 
in the case of $SU(2)$.

Since $(p,n)$ forms an isospin doublet of $SU(2)_V$, one may wonder 
whether another isospin doublet $(\Xi^0, \Xi^-)$ satisfies the same 
relation of vanishing coupling constants.  
However, under axial transformations of 
$SU(2)_R \times SU(2)_L$, 
$(\Xi^0, \Xi^-)$ transforms differently from $(p, n)$.  
In fact, we can verify that $\Xi$ components of $J^8$ and $J^3$ of 
(\ref{J38}) transform according to $(a = 1, 2, 3)$
\begin{eqnarray}
\, [ Q_{A}^{a}, J_{\Xi}^{8}] 
& = & \frac{1}{2} \gamma_5 \tau^a J_\Xi^{8} \, , \\
\, [ Q_{A}^{a}, J_{\Xi}^{3}] 
& = & - \frac{1}{2} \gamma_5 \tau^a J_\Xi^{3} \, .
\end{eqnarray}
These equations show that $J^8_\Xi$ and $J^3_\Xi$ 
carry the axial charge with opposite sign.  
The cross terms of 
(\ref{Pipn}) are now replaced by commutation relations, and therefore, 
$\pslash \gamma_5$ terms do not vanish~\cite{nemoto}.   
Thus for $(\Xi^0, \Xi^-)$, the coupling of, for example, 
$(\Xi^0)^* \to \Xi^0 + \pi^0$ does not vanish.  
There are two more ways to choose an $SU(2)$ subgroup out of 
$SU(3)$, as corresponding to the U and V spins.  
For the U spin, the two fundamental representations are 
$(\Sigma^-, \Xi^-)$ and $(p, \Sigma^+)$, and the adjoint
representation is $(K^0, \bar K^0, \bar d d - \bar s s)$.  
Thus, the coupling constant, for instance,  
for $(\Sigma^-)^* \to \Xi^- + K^0$ 
vanishes, while that of $p^* \to \Sigma^+ + K^0$ does not.  
Similarly, for the V spin, 
the fundamental and adjoint representations are 
$(\Sigma^+, \Xi^0)$, $(n, \Sigma^-)$ and 
$(K^+, K^-, \bar uu - \bar ss)$.  
Whether the properties of U and V spin symmetries 
are well realized in the real world or not 
depends on whether the relevant symmetry of the subgroup is good
or not. 

Finally, we would like to make a comment on the recent work of Kim and 
Lee~\cite{kimlee} from a point of view of chiral properties of a
fermion.  
For negative parity resonances, 
they have adopted an alternative interpolating field 
which involves a derivative, whose coupling to the negative parity 
resonance state is parameterized as 
\beq
\label{MEkimlee}
\langle 0 | J_{N^*}|N^*\rangle 
=
i \lambda_{N^*} \gamma_5 z_\mu \gamma^\mu u_{N^*} \, ,
\eeq
where $z_\mu$ is an auxiliary space-like vector which is orthogonal 
to the four momentum carried by the resonance state.   
The important point of (\ref{MEkimlee}), in contrast to 
our matrix element (\ref{JNtoneg}), is that 
$\gamma_\mu$ matrices change the chirality of the resonance state.  
Namely, the negative parity nucleon 
produced by $J_{N^*}$ of (\ref{MEkimlee}) has opposite chirality to 
that of the 
positive parity baryons produced by (\ref{eq:nucur}).  
Because of the different choice of interpolating fields,  
there is no relation between positive and negative parity baryons.  
Therefore,  similar results for coupling constants do not 
follow any more.  
It is emphasized, however, that their treatment of positive and 
negative parity baryons is 
quite different from the present one 
where they are regarded as parity doublet.

In summary, we have investigated positive and negative parity baryons 
($B$ and $B^*$) 
coupled by the same type of 
interpolating field, where $B$ and $B^*$
are considered to 
form a parity doublet.  
We found that the properties of $B$ and $B^*$ 
under the chiral transformation determine whether the meson-$BB^*$ 
coupling vanishes in the soft-meson limit, or not.  
We have shown  that 
the coupling vanishes if $B$ and $B^*$ carry the same axial 
charges.  For the $\pi NN^{*}$ coupling the $SU(2)$ triplet 
axial charges are indeed the same for $N$ and $N^*$, while for the 
$\eta NN^{*}$ the singlet 
axial charges are different for $N$ and $N^*$.
This leads to the 
suppression of the $\pi NN^{*}$ coupling while there is no 
such suppression for $\eta NN^{*}$.  
In the real world, the negative parity nucleon $N(1535)$ 
seems to satisfy these properties reasonably well.  
Finally, 
we have extended this argument to $SU(3)$ baryons and shown that 
which meson-$BB^*$ couplings vanish due to the same reasoning.  
It would be interesting if we could see experimentally
similar relations for various 
meson-$BB^*$ couplings.  

\vspace*{0.5cm}
\noindent
{\bf Acknowledgments}\\
This work is supported in part by the Grant-in-Aid for scientific 
research (A)(1) 08304024 and (c)(2) 08640356 and also by the
Grant-in-Aid for Encouragements of Young Scientists
of the Ministry of Education, Science and Culture of Japan.

\end{document}